\documentclass[%
 reprint,
 amsmath,amssymb,
 aps,
]{revtex4-2}
\usepackage{xcolor}
\usepackage{graphicx}% Include figure files
\usepackage{dcolumn}% Align table columns on decimal point
\usepackage{bm}% bold math
\usepackage{hyperref}% add hypertext capabilities
\usepackage[mathlines]{lineno}% Enable numbering of text and display math
\usepackage{lineno}
\newcommand{\aleksandra}[1]{{\color{black} {#1}}} %Aleksandra

\newcommand{\new}[1]{\textcolor{black}{#1}}

\begin{document}

\preprint{APS/123-QED}

%\title{Resolving Spinning Family Issues Among the Ancestors\\of GW231123-like Hierarchical Black Hole Mergers}% Force line breaks with \\
%\title{Resolving Spinning Family Issues Among the Ancestors\\of Very Massive Hierarchical Black Hole Mergers like GW231123}% Force line breaks with \\
%\title{Black Hole Family Issues: Tension among the Rapidly Spinning Ancestors\\of Very Massive Gravitational-Wave Events like GW231123}% Force line breaks with \\
\title{Resolving Black Hole Family Issues Among the Massive Ancestors\\of Very High-Spin Gravitational-Wave Events Like GW231123}% Force line breaks with \\
%\title{Resolving Spinning Family Issues Among the Black Hole Ancestors\\of Very Massive Gravitational-Wave Events like GW231123}% Force line breaks with \\

\author{Jakob Stegmann}
\email{jstegmann@mpa-garching.mpg.de}
\affiliation{%
Max Planck Institute for Astrophysics, Karl-Schwarzschild-Str.~1, 85748 Garching, Germany
}%

\author{Aleksandra Olejak}
\affiliation{%
Max Planck Institute for Astrophysics, Karl-Schwarzschild-Str.~1, 85748 Garching, Germany
}%

\author{Selma E.~de Mink}
\affiliation{%
Max Planck Institute for Astrophysics, Karl-Schwarzschild-Str.~1, 85748 Garching, Germany
}%

\date{\today}% It is always \today, today,

\begin{abstract}
{\color{black}  % Selma is fooling around here... 

The latest detection of GW231123, a black hole (BH) merger with exceptionally large component masses and high spins, has been suggested as a smoking gun for hierarchical formation. In this scenario, a first generation of BHs form from collapsing stars in dense environments such as star clusters, where they assemble dynamically and undergo subsequent mergers. 
We discuss three challenges for forming GW231123-like events in this scenario: (1) The high masses of the incoming BHs appear to be in the predicted pair-instability mass gap suggesting that higher-order generation BHs are involved. (2) Very high spins ($\chi_f \gtrsim 0.8$) are unlikely for dynamically assembled BHs because of the \textit{isotropic} spin distribution. (3) Hierarchically formed BHs can receive large recoils that kick them out of their cluster and prohibit subsequent mergers.  We simulate this scenario and show that only a few percent of mergers recover remnants within GW231123's primary spin estimate $\chi_1=0.9^{+0.10}_{-0.19}$ and are retained inside typical star clusters. 

A large fraction of very rapidly spinning second-generation BHs (including $\chi_f\gtrsim0.9$) can form if the first-generation BHs merges with \textit{aligned spins}. This is a natural outcome of massive binary star evolution scenarios, such as a chemically homogeneous evolution. This scenario also predicts equal masses for the components, implying that the resulting BHs receive low recoil kicks ($v_k\lesssim100\,\rm km\,s^{-1}$) and would therefore likely be retained inside a cluster. GW231123-like events, if formed in a star cluster, could require first-generation BHs with aligned spins formed from interacting stellar binaries, followed by the dynamical assembly for a subsequent merger.

}
\end{abstract}

\maketitle

%\tableofcontents

The LIGO–Virgo–KAGRA collaboration has reported the detection of GW231123, a gravitational-wave signal from the most massive binary black hole merger observed to date \citep{GW231123}. The two black holes have inferred masses of $m_1=137^{+22}_{-17}\,\rm M_\odot$ and $m_2=103^{+20}_{-52}\,\rm M_\odot$ and exhibit evidence for the large spins $\chi_1=0.9^{+0.10}_{-0.19}$ and $\chi_2=0.8^{+0.20}_{-0.51}$ seen so far. The exceptionally large masses and spins suggest that the two black holes themselves were build up by one or more preceding black hole mergers \citep{Gerosa2021}. First, this is because their masses are consistent with lying inside the putative upper mass gap at about $60$~--~$130\,\rm M_\odot$, where black holes are not expected to form through stellar collapse \citep{Bond1984,Heger2002,Fryer2001}. Second, large spins of about $0.6$~--~$0.8$ are naturally achieved through hierarchical mergers \citep{Campanelli2006,Berti2008,Gerosa2017,Ghersi2021}, whereas most merging black holes that presumably formed through stellar collapse were inferred to be slowly spinning \citep{GWTC3}.

Hierarchical mergers are expected in regions of large stellar densities such as star clusters \citep{Rodriguez2019,Antonini2019,Mapelli2021,AntoniniGieles2023,Chattopadhyay2023,Fragione2023,Mahapatra2025} or active galactic nuclei \citep{Yang2019,ArcaSedda2020,Vaccaro2024,Gilbaum2025}, where gravitational few-body interactions facilitate the dynamical assembly and mergers of multiple generation of binary black holes. As such, hierarchical mergers have been invoked to explain the properties of a set of previously detected binary black hole mergers \citep{Yang2019,Rodriguez2020,Mapelli2021,Mahapatra2021,Vigna2021,Alvarez2024,Mahapatra2024}. For GW231123 a hierarchical merger scenario in which both of its black holes were dynamically assembled is further supported by the inference of non-zero tilt angles between the black hole spins and the orbital angular momentum vector \citep{GW231123}. However, recovering the properties of the GW231123's black holes through hierarchical mergers is not trivial considering the following challenges.

\begin{enumerate}
    \item\label{item-mass} \textit{Upper mass gap:} Achieving the primary's mass $m_1\gtrsim120\,\rm M_\odot$ through hierarchical mergers of lighter black holes requires lifting the lower end of the canonical mass gap \citep{Heger2002,Farmer2020,Farag2022} or a \new{sufficiently dense environment that would enable a} sequence of mergers involving at least three black holes.
    \item\label{item-spin} \textit{Very large spin:} Hierarchical mergers of dynamically assembled black holes are thought to yield remnant spin distributions centred around $\chi_f\approx 0.7$ \citep{Fishbach2017,Gerosa2017}. This is consistent with the lower end of the uncertainty range $\chi_1=0.9^{+0.10}_{-0.19}$ of GW231123's primary and well within the secondary's range, but significantly below $\chi_{1,2}\gtrsim0.8$ which is strongly supported by the LVK analysis \citep{GW231123}.
    \item\label{item-retention} \textit{Cluster retention:} Hierarchically formed black holes can receive large recoil kicks due to the asymmetric loss of linear momentum by gravitational-wave emission during the merger \citep{Bonnor1961,Peres1962,Bekenstein1973,Fitchett1983}. The kick magnitude can be particularly large for unequal-mass mergers \citep{Gonzales2007} and unequal spins \citep{Lousto2012} and may exceed the escape speed of a dense stellar environment \citep{Antonini2016,Gerosa2019}. In particular, \citet[][Sec.~6.5]{GW231123} argue GW231123's properties would prefer the primary to be formed in a merger of black holes with masses $109^{+26}_{-32}\,\rm M_\odot$ and $35^{+35}_{-19}\,\rm M_\odot$ and the secondary from $78^{+28}_{-39}\,\rm M_\odot$ and $22^{+31}_{-15}\,\rm M_\odot$. The recoils of such unequal masses would exceed the typical escape velocities of young star clusters and globular clusters \citep{Antonini2016} and would require GW231123 to occur in environments with significantly higher escape velocities, \new{such as nuclear star clusters and active galactic nuclei, see e.g., \citet{Delfavero2025}}.
\end{enumerate}

\begin{figure}
    \centering
    \includegraphics[width=\linewidth]{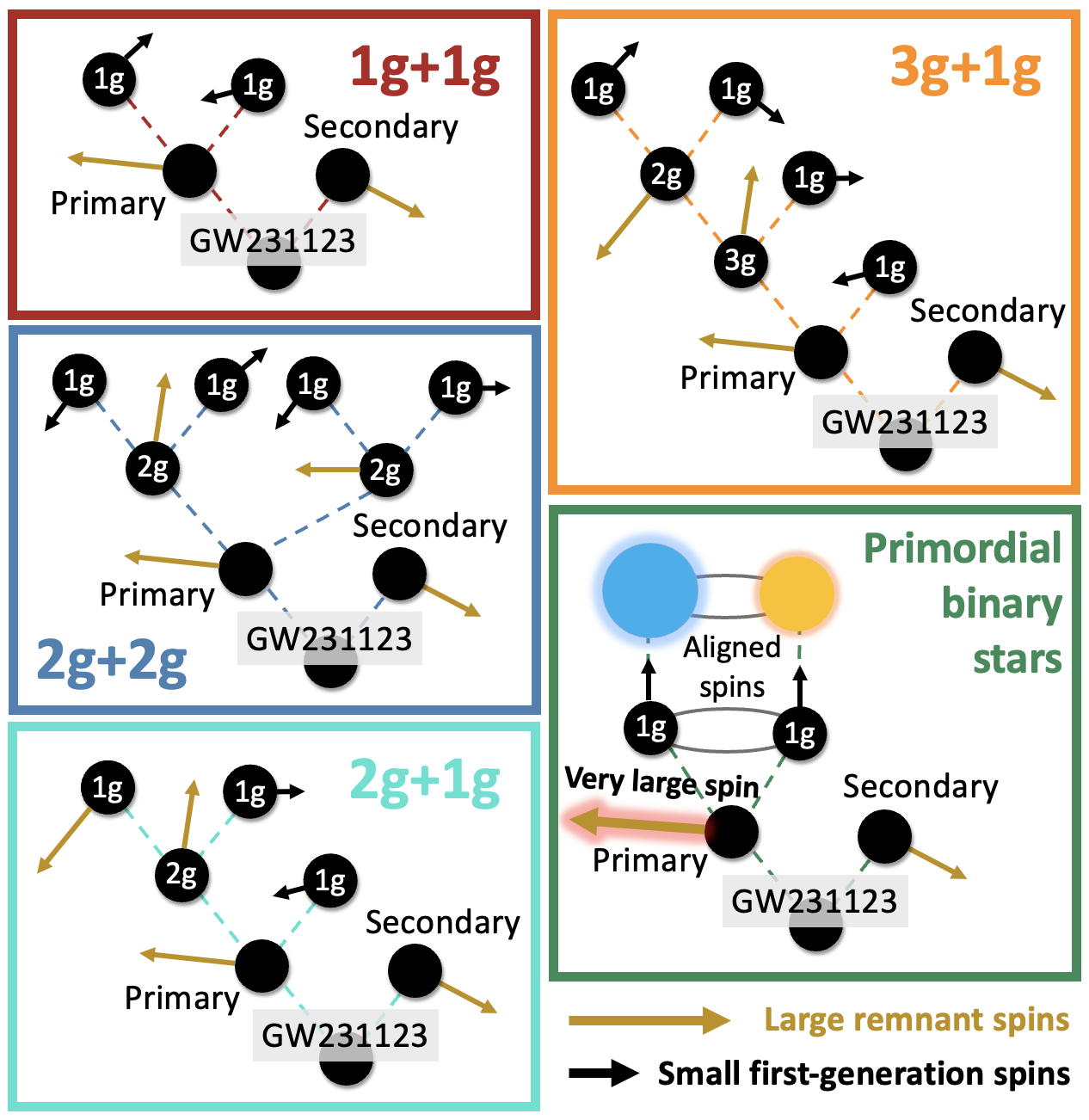}
    \caption{Cartoons of formation scenarios considered in this work. For purely dynamical scenarios we assume that the primary black hole is formed through a hierarchical sequence 1g+1g (red), 2g+2g (blue), 2g+1g (turquoise), and 3g+1g (orange), and subsequently merges with a secondary black hole in a GW231123-like event. For these dynamical formation sequences we assume the 1st-generation black holes (1g) to be slowly spinning \citep{GWTC3} the remnants to obtain a large spin. At each merger, the directions of the spin and orbital angular momentum vectors are fully randomised. Alternatively, we consider a scenario where the primary black hole forms from massive binary star evolution (green), e.g., of primordial stellar binaries in a star cluster, before it dynamically assembles and merges with the secondary black hole. Here, we assume the 1g black holes to obtain aligned spins (exploring different assumptions about their magnitudes) through binary star evolution which leads to much higher primary spins $\chi_f\approx0.9$.}
    \label{fig:sketch}
\end{figure}

\section{Hierarchical mergers of dynamically assembled binary black holes}\label{sec:isolated}

Focussing on the challenges \ref{item-mass}~--~\ref{item-retention}, we discuss how massive, very highly spinning merging black holes---such as those seen in GW231123---could be formed through a sequence of dynamically assembled black hole mergers. We begin to consider the ``canonical" hierarchical merger scenario in which they form dynamically through gravitational few-body encounters of single or binary black holes, as extensively studied for dense star clusters \citep{Downing2010,Samsing2014,Rodriguez2016,Antonini2016,Askar2017,Park2017,Rodriguez2018,Martinez2020,AntoniniGieles2020,Mapelli2021}. Our method is similar to that of \citet[][]{Fishbach2017}. We assume a first generation of $N=10^5$ merging black holes whose remnants participate in a sequence of hierarchical mergers (see sketch in Figure~\ref{fig:sketch}). We assume that the first generation of black holes is slowly spinning, determined from the black hole spin distribution inferred by \citet[][]{GWTC3} from previous gravitational-wave detections. Since in a dynamical assembly scenario black holes are paired from random directions the spin and orbital angular momentum directions of any merging black holes are mutually independent and isotropically distributed. For any merger, we compute the recoil kick velocity $v_k=v_k(q,\mathbf{\chi_1},\mathbf{\chi_2},\mathbf{\hat{L}})$ and remnant spin $\chi_f=\chi_f(q,\mathbf{\chi_1},\mathbf{\chi_2},\mathbf{\hat{L}})$ of the resulting black hole using Eqs.~from \citet[][]{Lousto2012} and \citet[][]{Hofmann2016}, respectively, where $0<q\leq1$ is the mass ratio, $\mathbf{\chi_{1,2}}$ are the black hole spin vectors, and $\mathbf{\hat{L}}$ is the direction of the orbital angular momentum vector. Remnants whose recoil kicks exceed a given ejection speed $v_{\rm ej}$ are removed from participating in further mergers. Otherwise, we use the remnant spin as input for computing the properties of succeeding mergers. The exact sampling procedures of our hierarchical merger sequences are detailed in Appendix~\ref{appendix-A}. In particular, we assume that any merger between two first-generation (1g+1g) and two second-generation black holes (2g+2g) is equal-mass ($q=m_2/m_1=1$), as preferred by cluster dynamics \citep{Rodriguez2019,Torniamenti2024}. For mergers involving black holes of different generations, we additionally consider merger sequences with constant first-generation black hole masses leading to $q=1/2$ for 2g+1g mergers and $q=1/3$ for 3g+1g mergers, respectively.

%we assume that any merger is between equal-mass black holes ($q=m_2/m_1=1$). On the one hand, this is preferred by gravitational dynamics of dense stellar environment where black holes of similar masses tend to merge due to mass segregation \citations. On the other hand, we find that---for a first generation of slowly spinning black holes---assuming $q<1$ impedes achieving low recoil kicks and large spins  \citations. Moreover, we may expect to form GW231123's black holes, which occupy the high-mass tail of the black hole mass distribution, most efficiently through the mergers of near-equal-mass mergers  \citations.

\begin{figure*}
    \centering
    \includegraphics[width=\linewidth]{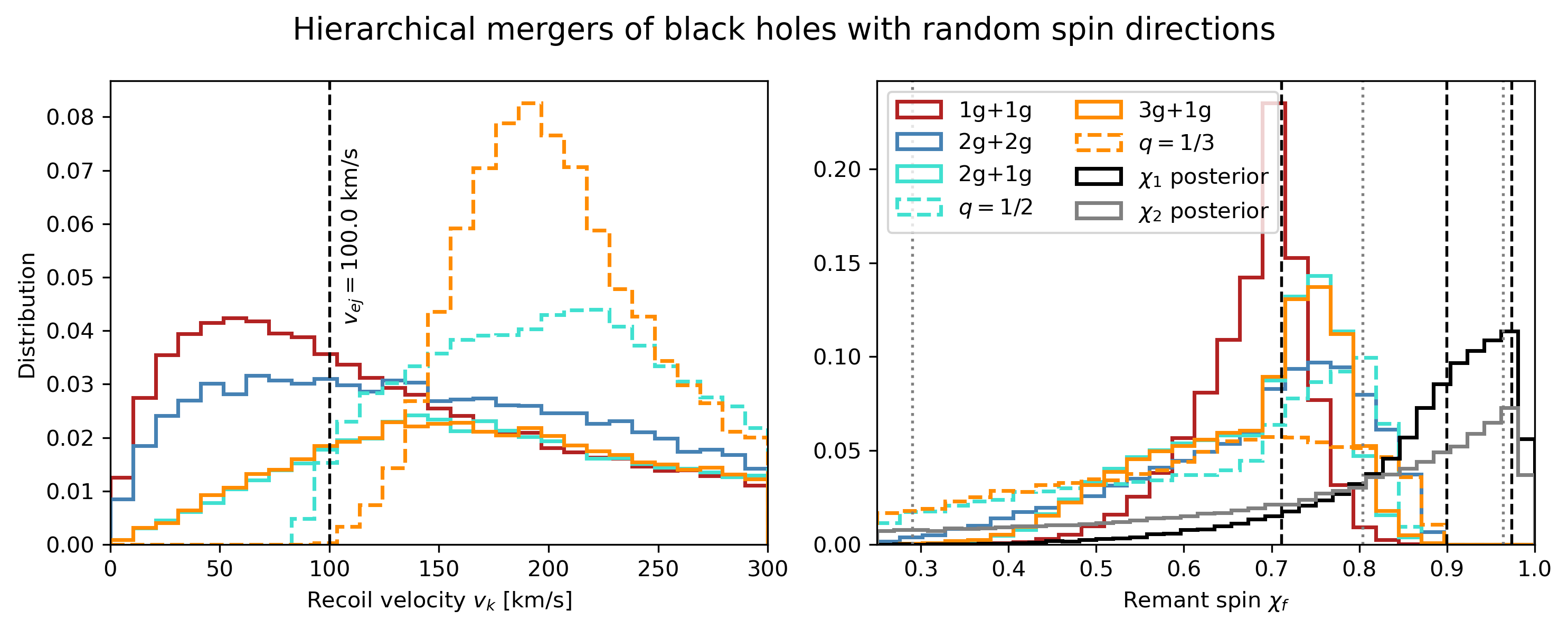}
    \caption{Recoil kick velocity distribution (left panel) and remnant spin distribution (right panel) of black holes formed through 1g+1g (red), 2g+1g (turquoise), 2g+2g (blue), and 3g+1g (orange) merger sequences. For all mergers, it is assumed that the black hole spin and orbital angular momentum directions are randomly distributed. For the coloured solid lines, we assume all mergers result from incoming black holes of the same mass. The dashed turquoise and orange line assume $q=1/2$ and $q=1/3$ for 2g+1g and 3g+1g, respectively. The vertical dashed line in the left panel indicates the adopted ejection velocity $v_{\rm ej}=100\,\rm km\,s^{-1}$, beyond which remnant black holes are removed from participating in further mergers (see text). In the right panel, we show GW231123's \texttt{Combined} posterior distributions for the primary (black) and secondary spin (grey) from the public LVK data release \citep{GW231123Zenodo}. Vertical dashed and dotted lines indicate their $0.1$, $0.5$, and $0.9$-quantiles, respectively.}
    \label{fig:isotropic}
\end{figure*}

Figure~\ref{fig:isotropic} shows the distributions of recoil kicks (left panel) and remnant spins (right panel) black holes obtained in a globular cluster scenario with $v_{\rm ej}=100\,\rm km\,s^{-1}$. In general, we observe that mergers between black holes with similar spin magnitude (1g+1g and 2g+2g) have a higher probability to yield small recoils below $v_{\rm ej}$. Meanwhile, the largest remnant spins are obtained if both preceding black holes themselves result from previous mergers (2g+2g) and are therefore highly spinning ($\chi_f\approx0.7$), whereas the smallest maximum spins are obtained when both black holes are slowly spinning (1g+1g). 
Comparing different formation sequences of equal-mass mergers, we find a fraction 8.1$\times10^{-2}$ of the 1g+1g mergers yield recoils $v_k<v_{\rm ej}= 100\rm\, km\,s^{-1}$ and remnant spins $\chi_f>0.7$ (the primary's 0.1-quantile \citep{GW231123Zenodo}),
1.1$\times10^{-1}$ for 2g+2g mergers,
3.9$\times10^{-2}$ for 2g+1g mergers,
and 3.8$\times10^{-2}$ for 3g+1g mergers. Considering our unequal-mass models, we find the general spin distribution to be slightly shifted to higher values. However, as a result of stronger recoil kicks only $1.1\times10^{-3}$ and none (out of $N=10^5$ systems) of the 2g+1g ($q=1/2$) mergers and 3g+1g ($q=1/3$) mergers, respectively, have $v_k<v_{\rm ej}= 100\rm\, km\,s^{-1}$ and $\chi_f>0.7$. Moreover, in
none of our models do mergers attain the remnant spins above the posterior's median $\chi_1\approx0.9$. 
The maximum fraction of mergers with remnant spins above a fiducial value $\chi_f=0.8$, which is strongly supported by all waveform models explored by \citet{GW231123}, and recoils below $v_{\rm ej}$ is 3.8$\times10^{-2}$ for the 2g+2g mergers. In contrast, nearly all remnant spins in all considered formation sequences are within the measured spin range of the secondary due to its long posterior tail which extends to low values $\chi_2\approx0.3$ \citep{GW231123Zenodo}. These findings are consistent with previous studies on equal-mass hierarchical mergers with a slowly spinning first generation of black holes \citep{Rodriguez2019,Borchers2025} .
Our results only marginally shift if we consider escape velocities up to $v_{\rm ej}=500\,\rm km\,s^{-1}$ for nuclear star clusters. 2g+2g mergers would remain the most favourable formation sequence in order to reproduce the highest fraction of retained black holes with large spins, but only a fraction $1.1\times10^{-5}$ exceeds the primary's median spin $\chi_f>\chi_1\approx0.9$.

\begin{figure*}
    \centering
    \includegraphics[width=\linewidth]{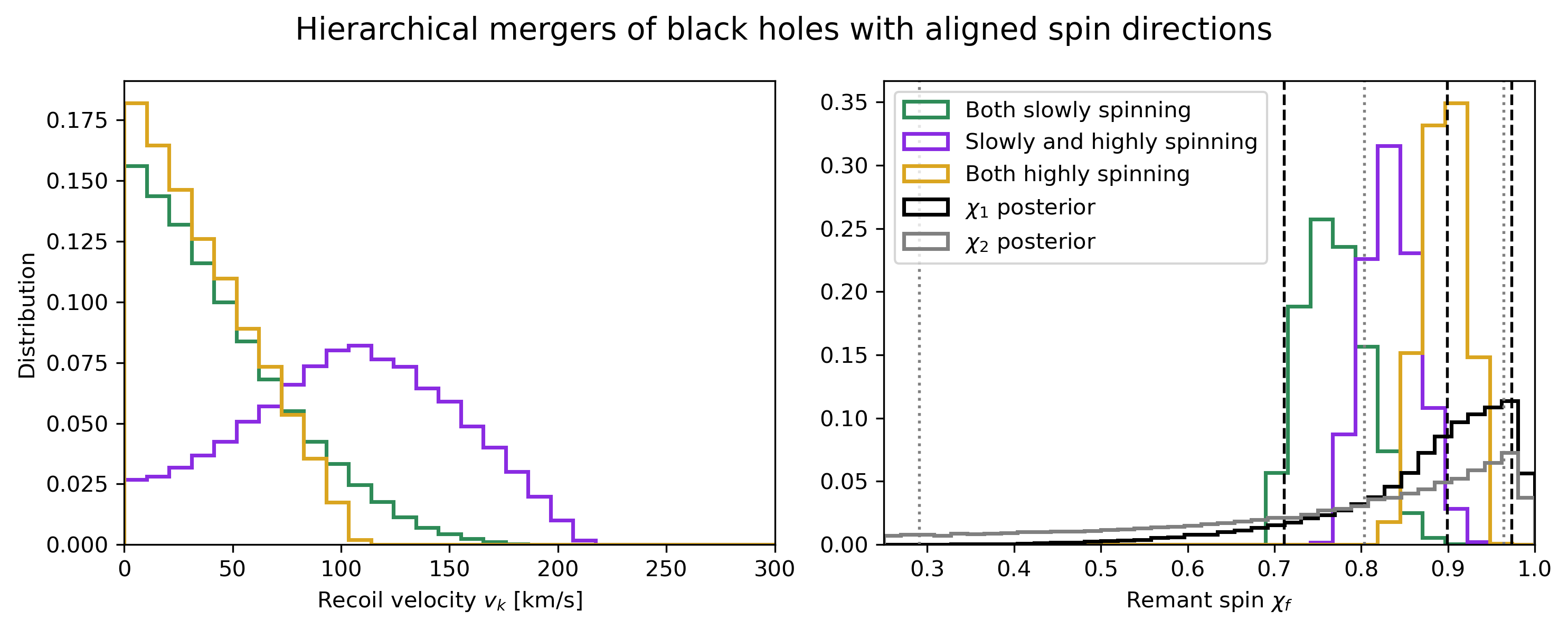}
    \caption{Same as Figure~\ref{fig:isotropic} for our primordial binary star scenario, where the ancestral black holes of GW231123's are assumed to have equal masses and  spins aligned with the orbital angular momentum. For the green line we assume both black holes to be slowly spinning, for the purple line we assume just one to be highly spinning, and for the gold line we assume both to be highly spinning (see text for details).}
    \label{fig:aligned}
\end{figure*}

Based on the remnant spin (\ref{item-spin}) and kick constraints (\ref{item-retention}), we find that neither of the above scenarios agree well with the observation but none of them could be confidently ruled out as formation pathways of GW231123's black holes due to large spin measurement uncertainties. In addition, all but the 1g+1g sequence could preserve the canonical mass gap of $60$~---~$130\,\rm M_\odot$ (\ref{item-mass}). Coincidentally, the 2g+2g sequence which yields remnant spins which are the closest to GW231123's primary spin posterior also recovers well its inferred mass range if involving four black holes with masses near the $\sim35\,\rm M_\odot$ peak \citep{GWTC3,Roy2025}. We stress that these considerations are independent of any rate calculation for each hierarchical formation pathway, which generally depend on the stellar environment \citep{Mandel2022} and, in practice, may be difficult to reliably obtain from simulations due to the expected low-number statistics of massive black holes (see \citet{Rodriguez2020} for a discussion on lower-mass binary black hole mergers). Any such rate calculation may further penalise the plausibility of a particular formation pathway, for instance, if cluster simulations would produce GW231123's primary at an insufficient rate through 2g+2g mergers \citep{Rodriguez2019}. However, the fact that for any given formation sequence the fraction of retained black holes with very large remnant spins is concerningly small holds regardless of the expected rate of the formation sequence. For instance, if we were certain that the primary has indeed a spin above the fiducial value $\chi_1\gtrsim0.8$ we would expect to have seen several tens of GW231123-like detections with lower spins if formed through 2g+2g mergers and even more in the alternative formation sequences.

\section{Hierarchical mergers of black holes from binary star evolution}
Motivated by the fact that canonical hierarchical mergers yield remnant spin distributions peaking well below GW231123's spin posteriors, we explore an alternative formation channel involving massive binary star evolution. Even if the tension between the explored remnant spin distributions and GW231123's spin posteriors appears not be large due to measurement uncertainties, this may become more relevant if additional GW231123-like mergers are detected in the future with stronger evidence for very large component spins ($\chi_f\gtrsim0.8$). 

\begin{figure*}
    \centering
    \includegraphics[width=\linewidth]{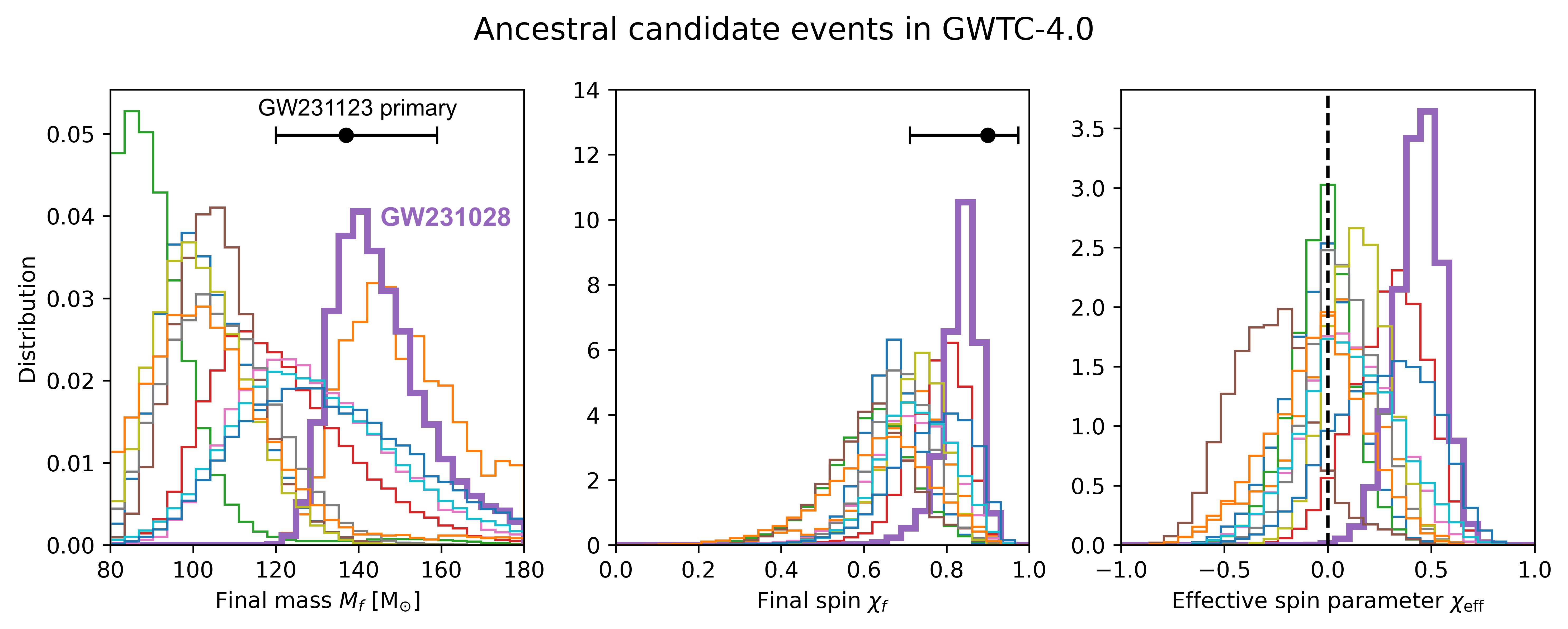}
    \caption{\new{Posterior samples of 13 massive binary black hole merger events (median $M_f>100\,\rm M_\odot$ and $M_f$ consistent with $m_1=137^{+22}_{-17}\,\rm M_\odot$) contained in the cumulative Gravitational-Wave Transient Catalog 4.0 \citep{GWTC4}. The posterior samples are retrieved from public data releases \citep{ligo_scientific_collaboration_and_virgo_2022_6513631,ligo_scientific_collaboration_and_virgo_2023_8177023,ligo_scientific_collaboration_and_virgo_2025_17014085}. Thick purple lines indicate the posteriors of GW231028, which is the only event with strong support for non-zero $\chi_{\rm eff}$, while error bars indicate the primary mass and spin uncertainties of GW231123, respectively.}}
    \label{fig:GWTC}
\end{figure*}

In the following, we consider a scenario in which very highly spinning massive black holes (such as GW231123's primary) originate from the evolution of massive primordial binary stars in stellar clusters, which then, dynamically assemble and merge with another black hole in a GW231123-like event (see Figure~\ref{fig:sketch}). The exact binary fraction of (massive) binary stars in clusters is uncertain and depend on the cluster properties \citep[e.g.,][]{Milone2012}. Observational surveys show evidence that the binary abundance in globular clusters may be somewhat smaller than in the field of Galactic disk and halo, but not by a large factor \citep{Hut1992}. While low-mass and wide binaries are likely to be disrupted by dynamical encounters, tight and massive binaries can remain bound \citep{Ivanova2005}. Observations also exhibit a preference for equal-mass binaries and show that binaries are more commonly found in cluster centres than single stars, as a result of mass segregation \citep{Milone2012, Giesers2019}. Recent observations suggest a higher fraction of binaries than predicted by simulations \citep{Giesers2019}; however, see also \citep{Muller2025,Bashi2025} who reports lower fraction of short-period binaries than expected. 

Effectively isolated massive binary stars are expected to produce merging black holes with spins aligned with the orbital angular momentum and large magnitude in some cases (see more in Sec. \ref{sec: binary_discuss}). Mergers of black holes with aligned, possibly large spins are known to yield remnant spins significantly above the nominal range of $\chi_f\approx0.6$~--~$0.8$ \citep{Campanelli2006,Alvarez2024}. Here, we explore the scenario by adopting three simple models, each assuming equal-mass black hole mergers with spins aligned to the orbit. Figure~\ref{fig:aligned} summarises the results. The first case, shown in green, considers both black holes to be slowly spinning, identical to the distribution of the first-generation black holes in Sec.~\ref{sec:isolated} (see Appendix~\ref{appendix-B} for details). The second case, shown in purple, assumes one slowly spinning black hole and one highly spinning black hole with a spin magnitude drawn from a uniform distribution $\mathcal{U}(0.5,1.0)$. The third case, marked in gold, considers both black holes to be highly spinning.

As a result, a large fraction of the resulting black holes---8.8$\times10^{-1}$ for both black holes slowly spinning, 4.7$\times10^{-1}$ for one slowly and one highly spinning, and 9.9$\times10^{-1}$ for both highly spinning---have recoil kicks below $v_{\rm ej}=100\,\rm km\,s^{-1}$ and remnant spins within the uncertainty range of GW231123's primary. The smaller fraction for one slowly and one highly spinning is due to the significantly higher recoil velocities and increases when considering stellar environments with larger escape velocities. We find the remnant spins of the retained black holes to distribute as $\chi_f=0.76^{+0.05}_{-0.04}$ (both slowly spinning),
$\chi_f=0.84^{+0.05}_{-0.04}$ (slowly and highly spinning),
and $\chi_f=0.90^{+0.03}_{-0.03}$ (both highly spinning), where uncertainties refer to the $0.1$ and $0.9$-quantiles. 

These findings agree well with the measurements in GW231123 and are robust against certain model variations. If we consider mass ratios as low as $q=2/3$, the fractions of retained black holes with consistent remnant spins are 4.5$\times10^{-1}$, 1.2$\times10^{-1}$, and 8.7$\times10^{-1}$, respectively. Drawing the high-spin components from a narrower distribution $\mathcal{U}(0.5,0.8)$ does \textit{increase} the fraction of retained black holes with consistent remnant spins to 6.4$\times10^{-1}$ (slowly and highly spinning) and $\approx1.00$ (both highly spinning), while only marginally shifting the remnant spin distribution to smaller values.

\new{The cumulative Gravitational-Wave Transient Catalog 4.0 \citep{GWTC4} contains multiple merger events which lead to the formation of GW231123-like primary black holes and appear to be consistent with our formation scenario. Specifically, of the 218 merger events reported so far there are 18 whose final remnant black hole mass ($M_f$) measurements overlap with GW231123's primary mass range. This number reduces to 13 if we only include events whose median final mass is $M_f>100\,\rm M_\odot$ which removes events with very large mass uncertainties. In Figure~\ref{fig:GWTC} we show their inferred posterior distributions of $M_f$, $\chi_f$, and $\chi_{\rm eff}$. Most of these events have broad posterior distributions consistent with $\chi_{\rm eff}\approx0$ ($90\,\%$ C.L.), none of them show unambiguous support for $\chi_{\rm eff}<0$, and one event (GW231028 \citep{GWTC4}) is strongly favouring a positive value $\chi_{\rm eff}=0.4^{+0.2}_{-0.2}$. Recently, \citet{Antonini2025} showed that the broadening of the  $\chi_{\rm eff}$ distribution of black hole mergers with masses above 40~--~$50\,\rm M_\odot$ is consistent with a subpopulation of mergers with large, isotropic spins from star clusters (see also \citet[][]{Tong2025} for a similar conclusion based on the sparsity of secondaries above those masses). We note that any massive merger with a clear support for negative $\chi_{\rm eff}$ detected in the future would further strongly support this scenario \citep{Rodriguez2018}. Here, we stress that the present sample which centres around $\chi_{\rm eff}\approx0$ is also consistent with some black hole spins being \textit{aligned} and ``both slowly spinning" or ``slowly and highly spinning", and may therefore consitute natural candidates to build up GW231123's primary from primordial binary stars in a stellar cluster. In particular, GW231028's support for positive $\chi_{\rm eff}$ strongly indicates that its black hole spins are co-aligned with the orbital angular momentum vector. Auspiciously, its final mass $M_f=144^{+27}_{-14}\,\rm M_\odot$ and final spin magnitude $\chi_f=0.84^{+0.05}_{-0.10}$ closely matches GW231123's primary mass and spin, respectively, and we find $\chi_1>0.5$ ($\chi_2>0.5$) in 57\,\% (52\,\%) of its posterior samples \citep{ligo_scientific_collaboration_and_virgo_2025_17014085}, which is consistent with the ``both highly spinning" model.}

We stress that the spin orientation of each black hole relative to the orbital angular momentum may change during the inspiral phase due to relativistic spin precession. As a result, the spin configuration established at large binary separations (for instance, when binary star evolution predicts nearly aligned spins) may not be preserved up to the moment of merger, which ultimately determines the remnant’s recoil and spin. However, for the equal-mass, aligned-spin systems we consider here, this deviation is expected to be negligible \citep[see discussion in][]{Hofmann2016}. Furthermore, dynamical interactions with randomly passing stars in dense stellar environments may both accelerate the merger and torque the binary orbit away from perfect spin–orbit alignment. This effect can become significant in very dense environments, such as nuclear star clusters \citep{Binney2008}, especially when the delay time between binary black hole formation and merger is long \citep[e.g.,][]{Binney2008}.

\section{Discussion of binary star evolution scenarios} \label{sec: binary_discuss}

A natural formation channel for binary black holes with aligned spins is isolated binary evolution \citep[see, e.g.,][]{Bavera2021b,Banerjee2023,Baibhav2024}. Moreover, very massive black holes, which are the subject of interest for our scenario, are expected to undergo so-called direct collapse, which imparts little to no natal kick \citep{Fryer12,Janka24,Baibhav2024}, leaving the spins aligned with the orbital angular momentum. Natal black hole spins predicted from isolated binary evolution are typically modest, with dimensionless spin parameters $\chi<0.4$ \citep{Bavera2021b,Belczynski2020b} consistent as inferred so far by \citet{GWTC3}. This outcome suggests efficient angular momentum transport within stars \citep{Spruit02,Fuller2019}, in line with constraints on stellar rotation profiles derived from asteroseismic measurements \citep{Cantiello2014,Eggenberger2022}.

However, several scenarios have been proposed in which one or both progenitor stars can be significantly spun up through binary interactions, ultimately producing highly spinning binary black hole mergers. For example, in the isolated common envelope or stable mass transfer channel, the progenitor of the second-born black hole can achieve high spin via tidal synchronisation in a tight black hole–Wolf–Rayet binary \citep{Qin2018,Bavera2021a,Olejak2021,Korb2025}. Moreover, chemically homogeneous evolution may naturally yield binary black hole mergers with large, nearly equal masses and high, strongly aligned spins \citep{deMink2016,Mandel2016, Marchant2016}. Another potential pathway to produce binary black holes with two rapidly spinning components is the double helium core scenario \citep{Qin2023} after, e.g., a double common envelope event \citep{Ivanova2013}.

A distribution centred on a final remnant spin above $\chi_f\gtrsim0.9$, as inferred for the GW231123 black hole components, favours a formation scenario involving perfectly aligned, equal mass, maximally spinning black holes with very high masses. Equal-mass, similarly spinning components also experience lower recoil kicks, increasing the likelihood that the remnant remains bound to its host cluster after the first merger. This makes the chemically homogeneous evolution scenario the most promising match for the explored scenario. Nevertheless, other scenarios involving aligned black hole spin components remain consistent with the posterior of components, potentially yielding distributions centred around $\chi_f\approx0.8$~--~$0.9$ (for systems with one highly spinning black hole) or $\chi_f\approx0.7$~--~$0.8$ (for systems where both black holes have lower spins).

The best observational candidates for highly spinning stellar-mass black holes come from X-ray binaries, where some black holes have estimated spin close to the critical value $\chi>0.95$ \citep[e.g.,][]{McClintock2006,Gou2009,Gou2011}. However, their measured spin values may be systematically overestimated due to several factors; see the recent review by \citet{Zdziarski2025}.

We highlight that the presented scenario requires the formation of stellar-origin black holes with masses above $\sim 60-80$ M$_{\odot}$ in order to reconcile the primary mass of GW231123, especially when accounting for the fact that a few percent of the total mass of the binary black holes are radiated away though gravitational waves. Such high masses were considered to lie within the so-called upper mass gap. In particular, pair-instability supernovae are expected to completely disrupt massive stars, leaving no remnant behind \citep{Bond1984,Heger2002, Fryer2001}. These processes were expected to produce a dearth of black holes within the mass range $\sim 50-135$ M$_{\odot}$\citep{Marchant2016, Mandel2016, Belczynski2016, Spera2017}. Recent studies, however, have shown that the exact location of the mass gap depends sensitively on the highly uncertain 12C($\alpha$,$\gamma$) 16O reaction rate, which can shift the edges of the gap to significantly lower or higher masses \citep{Farmer2020, Costa2021, Farag2022}. Adopting the lower limit within the 3$\sigma$-uncertainty of this reaction rate could potentially allow the formation of binary black hole mergers with component masses up to 90~--~100\,M$_{\odot}$ through isolated binary evolution \citep{Belczynski2020a, Olejak2022}. If the scenario presented here is realised in nature, it may have important implications for both the location of the upper mass gap and the underlying nuclear reaction rates. Alternatively, allowing for super-Eddington accretion during the second mass transfer phase could produce black hole masses above the pair-instability mass gap and result in a higher spin for one of the components \citep{vanSon2020,Briel2023}. \new{In contrast, this scenario is challenged by recent works on the latest gravitational-wave data \citep{GWTC4} by \citet[][]{Antonini2025} and \citet[][]{Tong2025} who interpret a broadening of $\chi_{\rm eff}$ for higher black hole masses and a dearth of massive secondary black holes, respectively, as the onset of a gap at $\gtrsim40$~--~$50\,\rm M_\odot$.}

\section{Summary}
The latest discovery of gravitational waves from the binary black hole merger GW231123 is exceptional not only because it involves the heaviest black hole masses $m_1=137^{+22}_{-17}\,\rm M_\odot$ and $m_2=103^{+20}_{-52}\,\rm M_\odot$ discovered so far, but also exhibits the strongest evidence for at least one very highly spinning component ($\chi_1=0.9^{+0.10}_{-0.19}$). Large masses and spins are thought to be a smoking gun for a hierarchical formation scenario in which the black holes themselves were the outcome of previous mergers. However, reconciling the strong evidence for a \textit{very large spin} of the primary $\chi_1\gtrsim0.8$ is difficult through the canonical hierarchical merger scenario, where the primary would originate from dynamically assembled black holes with random spin directions. In this work, we explored various hierarchical formation scenarios to form the two black holes observed in GW231123. Assuming mutually random spin directions, we find that only $\sim4$~--~$11\,\%$ of all 1g+1g, 2g+1g, 2g+2g, and 3g+1g mergers, respectively, yield remnant black holes which are retained in a typical cluster environment ($v_{\rm ej}<100\,\rm km\, s^{-1}$) and remnant spins consistent with the lower limit of the measurement uncertainty of GW231123's primary black hole. Auspiciously, we highlight recent work by \citet{Bamber2025} who obtained very large remnants spins through collisional simulations of black hole clusters in full general relativity, suggesting it $\sim0.9$ remnant spins to be more common. However, their specific choice of cluster properties and low-number statistics make it difficult to assess whether their cluster environment is a potential breeding ground for GW231123-like events.

Larger black hole spins and small recoil kicks may naturally be obtained if the ancestral black holes merged with spins aligned to the orbital angular momentum vector. For this case, we showed that most mergers yield remnant spins well within the highly spinning range of GW231123's primary, even if both incoming black holes are slowly spinning, and peak around $\chi_f\approx0.9$ near the primary spin median if both are highly spinning. Binary black hole mergers with aligned spins are expected from the evolution of massive binary stars. If placed in a dense stellar environment, their massive remnant black holes can dynamically pair with another black hole and give rise to a GW231123-like event. 

Among massive binary evolution channels, the chemically homogeneous evolution scenario may provide the most promising pathway for forming incoming black holes of GW231123-like events. Chemically homogeneous binary evolution is expected to produce strongly aligned, highly spinning, near-equally massive black holes. Moreover, the rather compact orbit of chemically homogenous binaries makes them less susceptible to perturbations from a cluster environment. We also find that scenarios where only one of the progenitor black holes is highly spinning, which would rather be a case in the stable mass transfer or common envelope formation channels, can still produce final spins approaching $\chi_f \approx 0.9$, significantly higher than typically expected from purely dynamical channels.

These findings can have important implications for our understanding of the upper black hole mass gap, implying that the mass gap could extend to $\sim 70$~--~$80\,\mathrm{M}_{\odot}$ or beyond. Such a shift remains compatible with existing uncertainties in stellar evolution models--particularly those associated with the poorly constrained $^{12}\mathrm{C}(\alpha,\gamma)^{16}\mathrm{O}$ reaction rate \citep{Farmer2020,Costa2021,Farag2022}.

\aleksandra{Later, several alternative scenarios have been proposed to explain the possible origin of GW231123. One class of models involves binary black hole mergers formed from isolated binaries of very massive stars above the upper pair-instability gap \citep{Croon2025, Popa2025, Gottlieb2025, Tanikawa2025}. In particular, high black hole spins and masses may arise from the chemically homogeneous evolution scenario \citep{Popa2025}. \cite{Gottlieb2025} further demonstrate that massive low-metallicity stars with moderate magnetic fields can form black holes consistent in mass and spin with GW231123. In particular, in their models collapse above the pair-instability gap assisted by rotation and magnetic fields can drive mass loss through disk winds and jet launching, enabling the formation of highly spinning black holes within the gap. 

Another alternative explanations invoke dynamical channels, however, under specific conditions and environments allowing for efficient accretion, such as active galactic nuclei disks \citep{Delfavero2025, Bartos2025} or dense clusters \citep{Kirouglu2025}. Accretion from a companion or the surrounding medium in these environments can further increase black hole masses and spins \citep{Bartos2025, Kirouglu2025}. Nuclear star clusters can retain higher-generation black hole merger remnants despite recoil kicks, enabling multiple generations of black hole mergers. \citet{Delfavero2025} found the observed masses and spins are most consistent with a merger between third- and fourth-generation black holes. Finally, some scenarios involving primordial black holes have also been suggested \citep{Yuan2025, Baumgarte2025, DeLuca2025}.
}

\section*{Acknowledgments}
We thank Jim Fuller, Mark Gieles, Fabio Antonini, Silvia A.~Popa, Davide Gerosa, and Stephen Justham for useful input and discussions.

\appendix

\section{Sampling procedures of dynamically formed mergers}\label{appendix-A}

\subsection{1g+1g}\label{sec:1g+1g}
We sample equal-mass binaries ($q=1$) with mutually independent isotropic spin and orbital momentum directions and spin magnitudes independently drawn from $\chi_{1,2}\sim{\rm Beta}(a=1.4,b=3.6)$ \citep{GWTC3}. We calculate the remnant recoil kick velocities and spins according to \citet[][]{Lousto2012} and \citet[][]{Hofmann2016}, respectively.

\subsection{2g+1g}\label{sec:2g+1g}
We sample remnant black holes of 1g+1g mergers and compute their recoil kick velocities and remnant spins as described in Appendix~\ref{sec:1g+1g}. If the computed kick velocities are below a given ejection velocity $v_k<v_{\rm ej}$ we let them merge with another equal-mass black hole whose spin magnitude is drawn from $\chi_{1,2}\sim{\rm Beta}(a=1.4,b=3.6)$. The spin directions are again assumed to be isotropically distributed and we calculate the final remnant recoil kick velocities and spins according to \citet[][]{Lousto2012} and \citet[][]{Hofmann2016}, respectively.

\subsection{2g+1g $(q=1/2)$}\label{sec:q=1/2}
We proceed as in Appendix~\ref{sec:2g+1g} but assume 2g black holes have twice the mass of 1g black holes.

\subsection{2g+2g}\label{sec:2g+2g}
We sample remnant black holes of 1g+1g mergers and compute their recoil kick velocities and remnant spins as described in Appendix~\ref{sec:1g+1g}. Remnant black holes with $v_k<v_{\rm ej}$ are again randomly merged with one another, assuming equal masses and isotropic spin directions. We calculate the final remnant recoil kick velocities and spins according to \citet[][]{Lousto2012} and \citet[][]{Hofmann2016}, respectively.

\subsection{3g+1g}\label{sec:3g+1g}
We sample remnant black holes of 2g+1g mergers and compute their recoil kick velocities and remnant spins as described in Appendix~\ref{sec:2g+1g}. If the computed kick velocities are below a given ejection velocity $v_k<v_{\rm ej}$ we let them merge with another equal-mass black hole whose spin magnitude is drawn from $\chi_{1,2}\sim{\rm Beta}(a=1.4,b=3.6)$. The spin directions are again assumed to be isotropically distributed and we calculate the final remnant recoil kick velocities and spins according to \citet[][]{Lousto2012} and \citet[][]{Hofmann2016}, respectively.

\subsection{3g+1g $(q=1/2)$}\label{sec:q=1/3}
We proceed as in Appendix~\ref{sec:3g+1g} but assume 3g (2g) black holes have thrice (twice) the mass of 1g black holes.

\section{Sampling procedures of mergers form binary star evolution}\label{appendix-B}

\subsection{Both slowly spinning}\label{sec:both-slowly}
We proceed as in Appendix~\ref{sec:1g+1g} but assume spin directions aligned with the orbital angular momentum vector.

\subsection{Slowly and highly spinning}
We proceed as in Appendix~\ref{sec:both-slowly} but draw one spin magnitude from a uniform distribution $\mathcal{U}(0.5,1.0)$.

\subsection{Both highly spinning}
We proceed as in Appendix~\ref{sec:both-slowly} but draw both spin magnitudes from a uniform distribution $\mathcal{U}(0.5,1.0)$.


\bibliography{apssamp}{}

% Produces the bibliography via BibTeX.

\end{document}